\documentclass[epj,final]{svjour}
\usepackage{latexsym}
\usepackage{graphics}

\begin{document}

\title{Pressure dependence of the Mg $3s4s^3S_1 \rightarrow 3s3p^3P_{0,1,2}$ 
transition in superfluid $^4$He}
\author{I. Baumann$^1$\and A. Breidenassel\and C. Z\"uhlke\and A. Kasimov\and
  G. zu Putlitz\and I. Reinhard \and K. Jungmann}
%
\institute{Physikalisches Institut der Universit\"at Heidelberg, Philosophenweg
  12, 69120 Heidelberg, Germany\\
$^1$Present address: The Boston Consulting Group GmbH \& Partner,
  Kronprinzenstr. 28, 70173 Stuttgart}
\mail{jungmann@physi.uni-heidelberg.de}
\date{Received: date / Revised version: date}
%
\abstract{The pressure shifts of the $3s4s^3S_1 \rightarrow 3s3p^3P_{0,1,2}$
  transition of magnesium atoms immersed in superfluid helium have been
  measured at $(1.3\pm0.1\,)$K between saturated vapour pressure and $24\,$bar. 
The wavelength is blue shifted
linearly by $(0.07\pm0.01)\,\frac{nm}{bar}$. This value can be satisfactorily
  described in the framework of the standard bubble model.
} 
\maketitle
\section{Introduction}
\label{intro}
Superfluid helium is a quantum substance with unique features, like the
phenomenon of superfluidity or the unusual dispersion curve
\cite{henshaw}. Despite a successful history and expanded research in this
field important properties of this quantum liquid remain still unexplained.\\  
Different experimental methods have been employed so far to study superfluid
helium. In general, they can be divided into two groups of conceptually 
distinguishable approaches. Firstly, the
superfluid itself is under investigation, which means parameters like its
density, its friction or its phase diagram are measured. Secondly, the interaction of
probe particles with the quantum fluid can be studied, e.g. the dispersion
curve has been measured with neutron scattering. This group comprises
experiments, where the experimental signal is derived from internal degrees of 
freedom of the microscopic probes. 
Foreign atoms and ions can be implanted and the
changes in their spectra reveal information about the interactions of the
probes with the helium environment \cite{toennies,weis,berndrev}. 
In the experiment reported here magnesium
atoms are introduced into the bulk superfluid and electronic transitions
within them are observed.\\ 

\begin{figure}
\begin{center}
\resizebox{0.4\textwidth}{!}{%
 \includegraphics{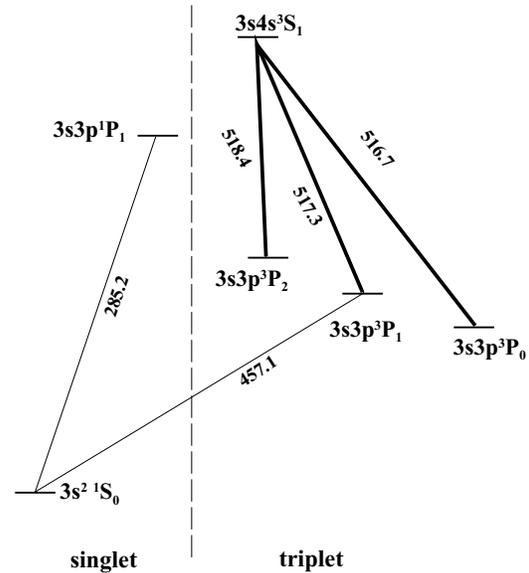}}
\end{center}
\caption{Energy level diagram of the triplet and singlet states of magnesium
  atoms. The numbers at the electronic transitions are the free atomic
  wavelength in nm.}
\label{fig:trans}       
\end{figure}
\noindent 
Foreign atoms or ions generally perturb the helium environment. 
Depending on the interaction between the probe particles and the superfluid
helium distinctly different defect structures are formed. If the density
around the foreign particle is lowered compared to the unperturbed helium
bulk a void with the foreign atom in its center forms; such structures are
known as bubbles. They are determined by
the interplay of repulsive and attractive interactions, the Pauli repulsion
between the electrons of the probe and the helium atoms surrounding it, as
well as the volume, respectively the surface
energy of the bubble. In contrast, there may be a strongly increased density,
even larger than the solidification density. These objects have been named
snowballs. They originate mainly from the very strong attractive polarization
forces between the foreign particle and the surrounding liquid and are
typically observed for positively charged particles, particularly for most of
the positive ions due to strong monopole induced dipole interactions 
\cite{atkins,toennies,weis,berndrev}.\\ 
From spectroscopy measurements \cite{mgpap} it is known that magnesium
atoms form bubble like structures under saturated vapour pressure. As a
surprising feature magnesium atoms show in liquid helium an unusual three times longer lifetime
for the $3s3p^3P_1 \rightarrow 3s^2\,^1S_0$ intercombination transition
compared to this transition in vacuum \cite{mgpap}. For other systems such a
behaviour has not been as pronounced as in this case. 
Therefore atomic magnesium
has been chosen to study the influence of an increased helium pressure
on a bubble-like structure in order to investigate whether this object is
stable at higher helium pressures and may even undergo observable 
structure changes.\\
Due to the interaction of the magnesium atoms with the surrounding superfluid
its electronic states are perturbed and the emission as well as the 
absorption lines of corresponding electronic transitions are shifted with
respect to their vacuum values. Further they are broadened and have asymmetric 
shapes \cite{berndrev}. The wavelength of the electronic transitions and the mean bubble
size can be predicted in the framework of a straightforward theoretical
approach, the standard bubble model. This is based on macroscopic quantities
such as surface and volume energies \cite{bubb1}. 
It has been successfully applied to singlet states so far exclusively
\cite{berndrev}. 
Here it is employed to describe triplet states as well.

\section{Experimental set-up}
\label{sec:setup}
\begin{figure}
\begin{center}
\resizebox{0.5\textwidth}{!}{%
\includegraphics{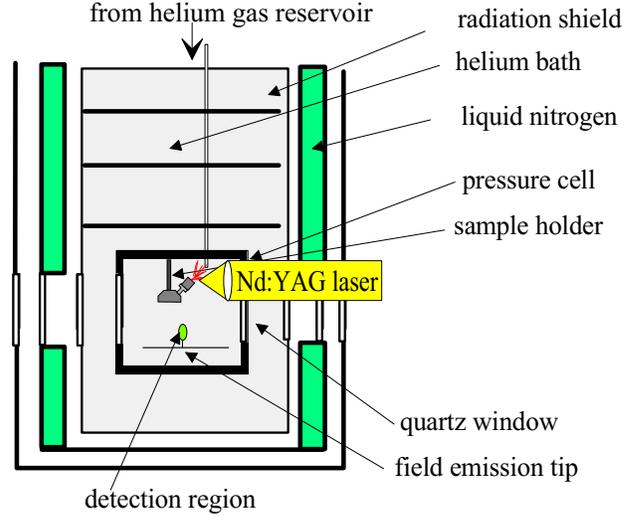}}
\end{center}
\caption{Cross section
  of the lower part of the helium bath cryostat. The pressure cell
  is mounted inside the liquid helium bath. Magnesium ions are produced by
  laser ablation, drawn by an electric field towards the bottom of the cell,
  where they recombine with electrons released from a field emission tip.}
\label{fig:cell}       
\end{figure}

A copper pressure cell (inner volume $=600\,$cm$^3$) is mounted inside a helium
bath cryostat (see figure \ref{fig:cell}). 
Its temperature is maintained between $1.2$ and $1.4\,$K. 
The cell is connected with a helium gas reservoir via a
capillary system (inner diameter $=1.5\,$mm) to allow filling by condensation
of helium gas. The liquid pressure can be adjusted by applying a corresponding 
helium pressure from the gas reservoir. Optical access to the cell is possible through 
three quartz windows (diameter $=39\,$mm) which are sealed by indium gaskets up
to $40\,$bar helium pressure at $1.2\,$K. \\ 
\begin{figure}
\begin{center}
\resizebox{0.5\textwidth}{!}{%
 \includegraphics{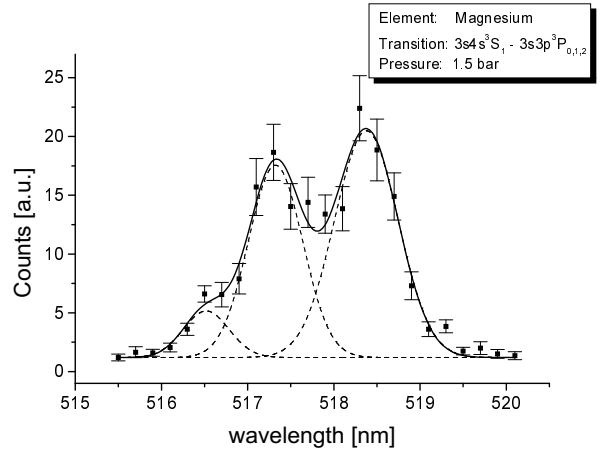}}
\end{center}
\caption{Recombination spectrum of the $3s4s^3S_1 \rightarrow 3s3p^3P_{0,1,2}$
  transition measured at an increased helium pressure of $1.5\,$bar.}
\label{fig:spek}       
\end{figure}
\noindent In the experiment the sample material under investigation has
a typical size of $5$x$5$x$5\,$mm. Ions are produced by laser ablation from the
surface of the sample with a focused Nd:YAG laser (focal diameter
$4.4\,\mu$m).  The laser energy is 
$8\,$mJ per pulse with a pulse width of $6-8\,$ns at wavelength 
$1064\,$nm \cite{inapap}. The ions  are drawn by
an electric field towards the bottom of the experimental chamber, where most of
them recombine with electrons from a field emission tip. 
The tip voltage was varied
between $-0.9$ and $-2.8\,$kV and the probe voltage between $0.6$ and
$1.2\,$kV. These voltages 
were adjusted for each pressure to maximize the signal to
noise ratio. These parameters correspond to electric fields between $0.4$ and
$1.0\,$ $\frac{kV}{cm}$ for a drift length of $42\,$mm.
The light emitted from the electron cascade after recombination
is imaged onto the entrance of a 
grating monochromator (Czerny-Turner type) with a wavelength resolution of
$0.025\,$nm. A photomultiplier tube (EMI S 20 extended) serves as detector, 
the signal
of which is digitized and recorded time resolved in $400$ bins of a width of $1.0\,$ms.\\
The recombination method as well
as the implantation and production of ions directly in the liquid based on the use
of laser ablation are both well established techniques \cite{berndrev}. 
In this experiment they were combined for the first time. 
Experimental data were taken at pressures in the full accessible pressure  range of the 
experimental method up to $24\,$bars,
where close to the solidification point 
the ion mobility drops dramatically with increasing pressure.\\
A typical spectrum of the
$3s4s^3S_1 \rightarrow 3s3p^3P_{0,1,2}$ transition at a
helium pressure of $1.5\,$bar is displayed in figure \ref{fig:spek}. The mean
wavelength of the three emission lines can be obtained by a fit of three
convoluted Gaussian line shapes.
\section{Calculation of the emission wavelength with the
  standard bubble model}
\label{chap:theo}
\begin{figure}
\begin{center}
\resizebox{0.5\textwidth}{!}{%
 \includegraphics{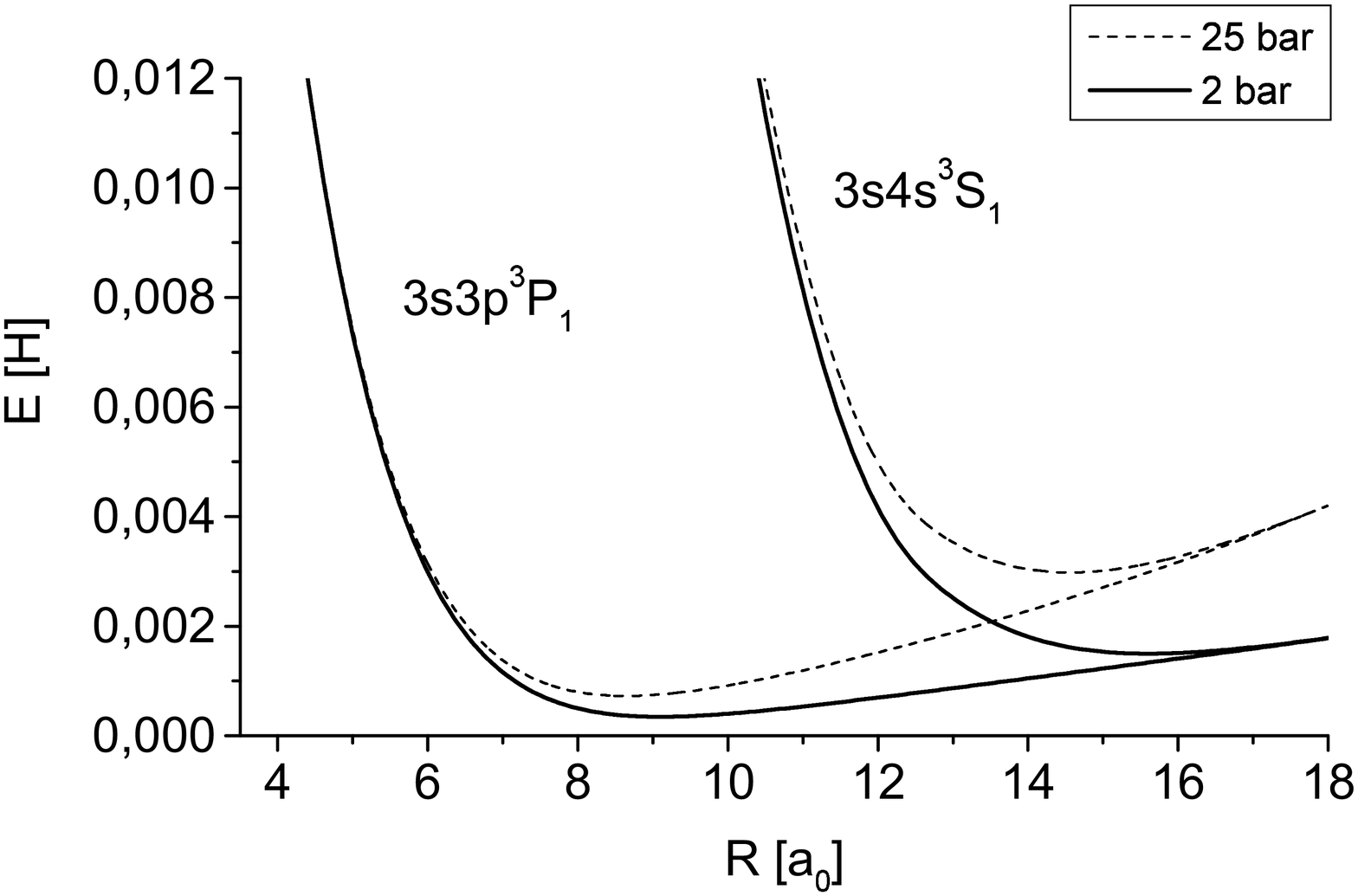}}
\end{center}
\caption{Defect energy of the triplet states $3s4s^3S_1$ and $3s3p^3P_1$ of
  the magnesium atom as a function of the internuclear 
magnesium-helium-distance R in units of hartree ($1\,$H$=27.212\,$eV)  
calculated by use of the bubble model (with $\alpha=1.18\,a_0^{-1}$,
$a_0=0.529\cdot10^{-10}\,$m). The dashed line corresponds to a pressure of
  $25\,$bar, the other one to a pressure of $2\,$bar.} 
\label{fig:f3}       
\end{figure}
\begin{figure}
\begin{center}
\resizebox{0.5\textwidth}{!}{%
 \includegraphics{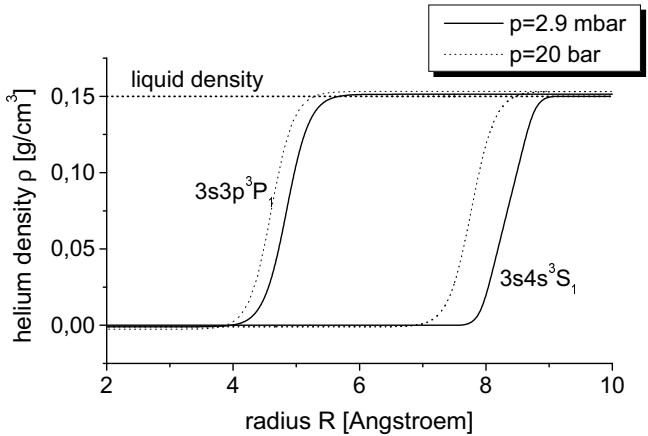}}
\end{center}
\caption{Density distribution of the helium environment around a magnesium
  atom in the $3s3p^3P_1$ state or in the $3s4s^3S_1$ state at 
  helium pressures of $2.9\,$mbar and
  $20\,$bar.} 
\label{fig:mgrad}
\end{figure}
The bubble model allows a prediction of the bubble size as well as of the
energy shift of electronic transitions compared with the free atomic case.\\
The total energy of the defect $E_{tot}$ is the sum of two terms, the
electronic contribution of the free atom $E_{free}$ and the so called defect energy
$E_{defect}$ \cite{bubb1}:
\begin{equation}
\label{eq:def}
E_{tot} = E_{defect}+E_{free}=E_{bubble}+E_{int}+E_{free}
\end{equation}
The defect part includes the bubble energy $E_{bubble}$ which is needed to
form the void and the pairwise interaction $E_{int}$ between the defect atom and surrounding
helium atoms. The bubble energy consists of macroscopic terms like volume
$E_{vol}$, surface $E_{surf}$ and volume kinetic $E_{vk}$ energies where the
later is due to the helium density gradient at the bubble surface \cite{bubb1,bubb2}:
\begin{eqnarray}
E_{bubble} & = & E_{vol} + E_{surf} + E_{vk}\\
& = &  \frac{4 \pi}{3}p R_B^3 \left( R_0 \right) + 4 \pi \sigma R_B^2 \left
  ( R_0\right) + \nonumber\\
& + & \frac{\hbar^2}{8m_{He}} \int_0^{\infty} 
\frac{(\bigtriangledown \rho (r,R_0, \alpha))^2}{\rho (r,R_0, \alpha )}\,d^3r
\end{eqnarray}
with the helium pressure $p$, the equilibrium bubble radius $R_B$, the radius
$R_0$ where the liquid density approaches zero, the width of the transition
region from the bubble to the helium environment $1/\alpha$, the surface
density $\sigma$ and the density $\rho(r,R_0,\alpha)$. The density follows an assumed
parametrization 
 \cite{bubb2}:
\begin{eqnarray}
\label{eq:rho}
\rho =\left\{ 
\begin{array}{l@{\quad}l}
0 & r<R_0\\
\rho_0\left[1-\left[1+\alpha\left(r-R_0\right)\right]e^{-\alpha\left(r-R_0\right)}\right] & r \ge R_0
\end{array} \right.
\end{eqnarray}
with the constant helium density $\rho_0=0.146\frac{g}{cm^3}$. This ansatz
assumes that helium is incompressible as $\rho(r,R_0,\alpha)$ can't be larger
than $\rho_0$.  
The bubble model has been successfully applied to describe experiments at elevated
helium pressures, e.g. for
electron bubbles the pressure dependence of electronic transitions can be very well calculated 
\cite{ebubb}. 
Further, there is a less than $20\,$\% change in  $\rho_0$ \cite{Donn} 
over the whole pressure range covered in this experiment and the associated 
relative difference in the calculated pressure shift, which arises from 
the last term in eq.(3), is below $2\cdot10^{-3}$.  
Therefore we find this assumption
motivated in our case.\\  
The defect energy is
obtained by adding the interaction energy $E_{int}$ of the states involved 
and the bubble energy. Multi particle
interactions are neglected in this approach and only pairwise magnesium- helium interactions are taken
into account \cite{beau}:
\begin{eqnarray}
E_{int}\left(S\right) & = & 4 \pi \int\limits_0^{\infty} V_{S}\left(r\right)
\rho\left(r,R_0, \alpha\right)r^2\,dr\\
E_{int}\left(P\right) & = & 4 \pi \int\limits_0^\pi sin \theta\,d\theta 
\int\limits_0^{\infty} [\left(cos \theta\right)^2V_{P}^\sigma
\left(r\right)+ \nonumber\\
& + & \left(sin\theta\right)^2V_{P}^\pi\left(r\right)] \rho\left(r,R_0,\alpha\right)\,r^2ß,dr
\end{eqnarray} \cite{berndrev}
with the interatomic pair potentials $ V_{S}$, $V_{P}^\sigma$, $V_{P}^\pi$,
where S stands for s-states and P for p-states, which are in the case of magnesium atoms in triplet
p-states only known without fine structure splitting \cite{pair}. The fine
structure splitting arising from spin orbit interactions is assumed not to
depend on the externally applied helium pressure, therefore a prediction for all
three emission lines can be made.\\
As the energy of the free atom is only an additive
contribution to the total energy, it can be neglected for the calculation of the radial dependence of
the defect energy, but has to be added for the calculation of the wavelength
of the electronic transitions. An example of the calculated defect energies
 of the two interesting
states $3s4s^3S_1$ and $3s3p^3P_1$  for two different helium pressures
($2.9\,$, $25\,$bar) is shown in figure \ref{fig:f3}.\\ 
The radius at the minimum
of the defect energy is the mean equilibrium radius of the defect structure in
the specific state. It
decreases with increasing pressure (see figure \ref{fig:mgrad}) for the $3s4s^3S_1$ state from $8.34\,$\AA\, at
$2.9\,$mbar to $7.68\,$\AA\, at $25\,$bar and for the $3s3p^3P_1$ state from
$4.85\,$\AA\, to $4.56\,$\AA\,. 
This decrease in the equilibrium radius with increasing pressure is qualitatively similar
to the behaviour of an electron bubble at an enhanced helium
pressure \cite{ebubb}. Additionally the model predicts the width of the transition
region from the bubble to the helium environment to be $0.45\,$\AA\,.\\
The sum of the free energy and the difference of the two defect state energies
yields a prediction of the
 pressure dependent emission wavelength of the transition $3s4s^3S_1 \rightarrow 3s3p^3P_1$:
 \begin{equation}
\lambda (p) = (516.48\pm0.01)\,nm - (0.08\pm 0.01)\,\frac{nm}{bar} \cdot
p\,[bar]
\end{equation}
The wavelength of the other two emission lines is obtained by adding the
respective fine structure splitting ( $+1.1\,$nm for $^3P_0$, $-0.53\,$nm for
$^3P_2$) to the zero pressure wavelength of $516.48\,$nm. 
\begin{figure}
\begin{center}
\resizebox{0.5\textwidth}{!}{%
 \includegraphics{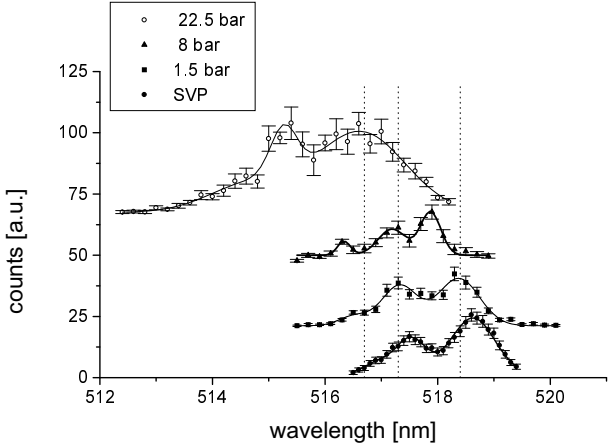}}
\end{center}
\caption{Emission spectra of the $3s4s^3S_1 \rightarrow 3s3p^3P_{0,1,2}$
  transition at different helium pressures ($3\,$mbar, $1.5$, $8$ and
  $22\,$bar). The dashed lines correspond to the free atomic
  transitions.} 
\label{fig:pspek}
\end{figure}
\begin{figure}
\begin{center}
\resizebox{0.45\textwidth}{!}{%
 \includegraphics{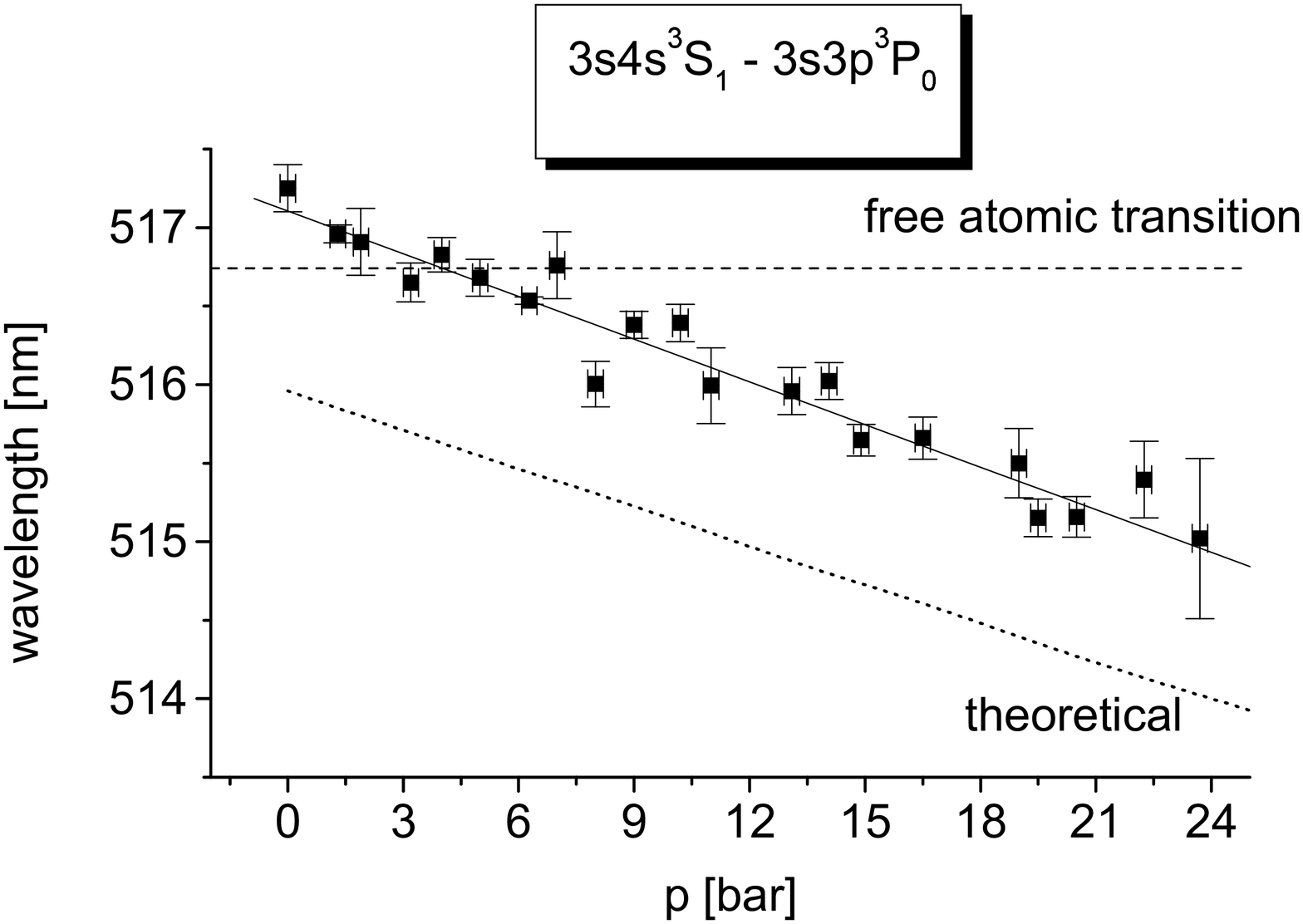}}
\end{center}
\caption{Emission wavelength of the $3s4s^3S_1 \rightarrow 3s3p^3P_0$
  transition of the magnesium atom as a function of the helium pressure. The
  dotted line corresponds to the emission wavelength calculated by use of the
  bubble model. The dashed line corresponds to the free atomic
  transition at $516.74\,$nm.} 
\label{fig:mp0}
\end{figure}
\begin{figure}
\begin{center}
\resizebox{0.45\textwidth}{!}{%
 \includegraphics{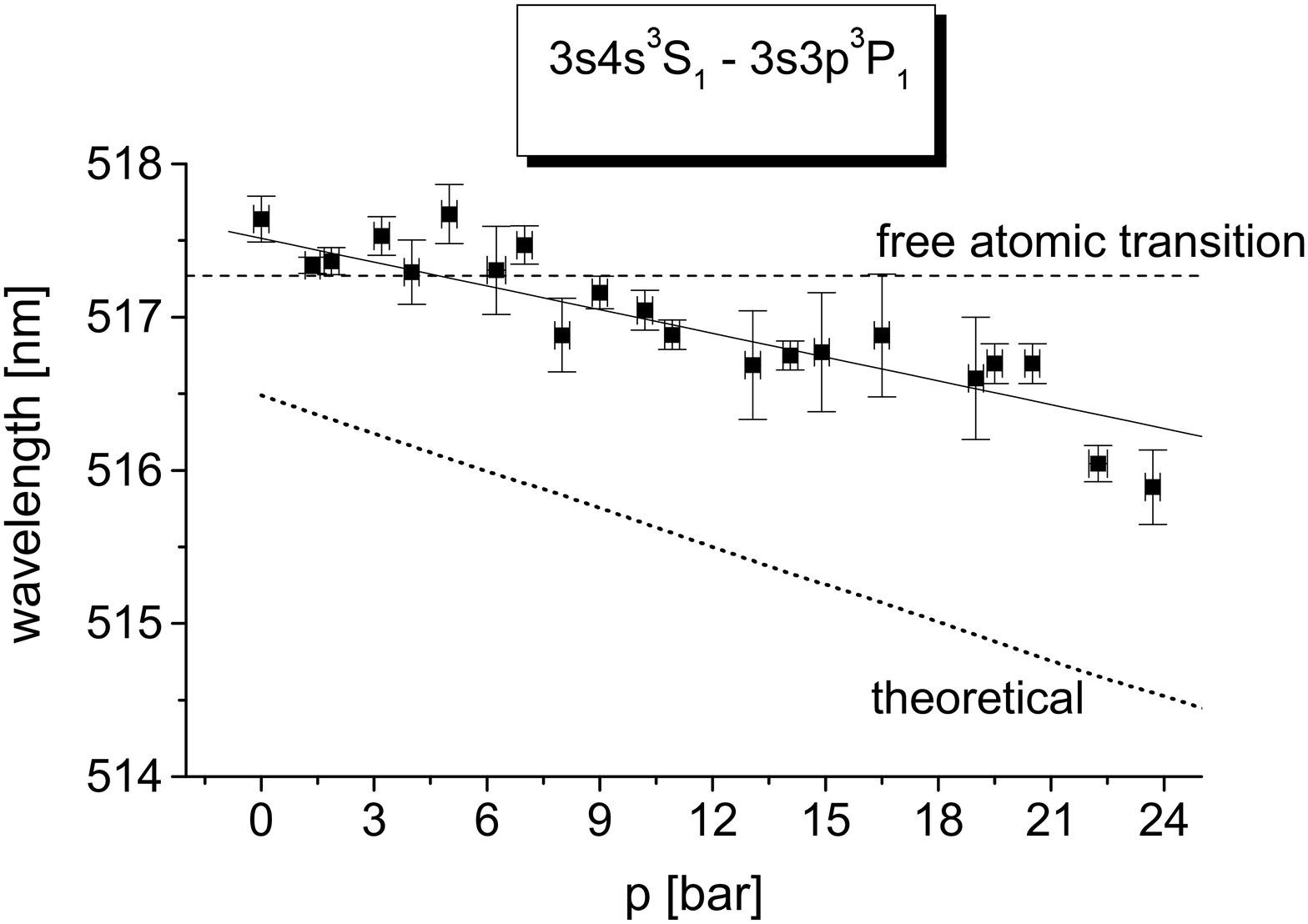}}
\end{center}
\caption{Emission wavelength of the $3s4s^3S_1 \rightarrow 3s3p^3P_1$
  transition of the magnesium atom as a function of the helium pressure. The
  dotted line corresponds to the emission wavelength calculated by use of the
  bubble model. The dashed line corresponds to the free atomic
  transition at $517.27\,$nm.} 
\label{fig:mp1}
\end{figure}
\begin{figure}
\begin{center}
\resizebox{0.45\textwidth}{!}{%
 \includegraphics{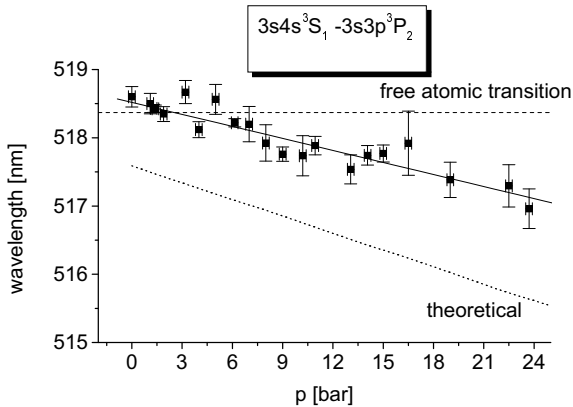}}
\end{center}
\caption{Emission wavelength of the $3s4s^3S_1 \rightarrow 3s3p^3P_2$
  transition of the magnesium atom as a function of the helium pressure. The
  dotted line corresponds to the wavelength calculated by use of the
  bubble model. The dashed line corresponds to the free atomic
  transition at $518.37\,$nm.} 
\label{fig:mp2}
\end{figure}
\section{Experimental results}
\label{sec:res}
Typical measured emission spectra for helium pressures $3\,$mbar,
$1.5\,$, $8\,$ and $22\,$bar are shown in figure \ref{fig:pspek}. 
The values below $1\,$bar were measured with another experimental cell
\cite{inapap} as the pressure cell allows measurements only at helium
pressures above $1\,$bar.\\  
The spectra  shift with increasing pressure to smaller wavelength, in
accordance with the
bubble model. The central emission wavelength of the three transitions is given
 in figure \ref{fig:mp0}, \ref{fig:mp1} and \ref{fig:mp2} as a function of 
the applied helium pressure.
 The error bars result from the line shape fits.  
The uncertainty of the wavelength calibration of the monochromator is
0.1~nm common to all points.
 The dotted line is the
calculated wavelength predicted by the standard bubble model (see chapter \ref{chap:theo}).\\
\begin{table*}[t]
\begin{center}
\begin{tabular}{|l|c|c||c|c|c|c|c|}
\hline
& transition & free & \multicolumn{2}{|c|}{superfl. helium} & \multicolumn{2}{|c|}{\rule[-3mm]{0cm}{1cm} superfl. helium} & ref.\\
& & atom & \multicolumn{2}{|c|}{saturated vapor pressure} & \multicolumn{2}{|c|}{increased  pressure} &\\
& & [nm] & [nm] & ref. & [$\frac{nm}{bar}$] & [$\frac{\%}{bar}$] & \\
\hline
e$^{-}$ & 1s-2p & & 11270 & \cite{ebubb}& 61 & 0.541 & \\
& 1s-1p & & 2480 & \cite{ref:grim92}& 252 & 10.161 
& \raisebox{1.5ex}[-1.5ex]{\cite{ebubb},\cite{ref:grim92}}\\
& 1s-1p & & 2480 & \cite{ref:grim92}& 300 & 12.097 & \cite{ref:golov95}\\
\hline
He$_2$ & $2^3S \rightarrow 2^3P$ & & 1083.2 & & -0.11 & -0.010& \\
& $2^3P \rightarrow 2^3S$ & \raisebox{1.5ex}[-1.5ex]{1083} & 1091.1 & \raisebox{1.5ex}[-1.5ex]{\cite{ref:hill}}& -0.3 & -0.027 & \raisebox{1.5ex}[-1.5ex]{\cite{ref:soley}}\\
\hline
Rb & $5^2S_{1/2} \rightarrow 5^2P_{1/2}$ & 794.76&  777.96 & \cite{ref:taka93}& 
-0.26 & -0.033 & \cite{ref:kino951}\\
\hline
Ba & $6s^2{}^1S_0 \rightarrow 6s6p^1P_1$ & 553.55 & 547.05 & \cite{ref:bauer901}
& -0.11 & -0.020 & \cite{ref:kan951}\\
\hline
Cs & $6^2P_{1/2} \rightarrow 6^2S_{1/2}$ & & 875.95 & & -0.26 & -0.030 & \\
& $6^2S_{1/2} \rightarrow 6^2P_{1/2}$ & \raisebox{1.5ex}[-1.5ex]{894.35} 
& 892.25 & \raisebox{1.5ex}[-1.5ex]{\cite{ref:taka93}}& -0.67 & -0.075 & \raisebox{1.5ex}[-1.5ex]{\cite{ref:kino952}}\\
\hline
Tm & $4f^{12} \left(^3H_5 \right)5d_{5/2}$ & 590.11& 596.21 & & & &\\
& $6s^2\left(5,5/2 \right)_{7/2}$ or & & & && &\\
& $4f^{13}\left(^2F^0_{7/2}\right)6s6p\left(^3P^0_1\right)\left(7/2,1\right)_J$ 
& 589.73 & 596.21 & & & & \\
& $\rightarrow4f^{13}\left(^2F^0_{5/2}\right)6s^2$ & & & 
\raisebox{4.5ex}[-4.5ex]{\cite{ref:ishika}}& 
\raisebox{4.5ex}[-4.5ex]{-0.06}& \raisebox{4.5ex}[-4.5ex]{-0.01}
&\raisebox{4.5ex}[-4.5ex]{\cite{ref:ishika}}\\
\hline
Mg & $3s4s^3S_1 \rightarrow 3s3p^3P_0$ & 516.73 &517.11 & this
& $-\left(0.09\pm0.01\right) $& -0.017 &this \\
& $3s4s^3S_1 \rightarrow 3s3p^3P_1$ & 517.27 & 517.51 & work 
& $-\left(0.06 \pm 0.01\right)$& -0.012 &work\\
& $3s4s^3S_1 \rightarrow 3s3p^3P_2$ & 518.36 & 518.52 && 
$-\left(0.06 \pm 0.01\right)$& -0.012 &\\
\hline
\end{tabular}
\end{center}
\caption{\label{tab:over}Electronic transitions of various elements measured
  at an increased helium pressure. The table includes the free atomic
  transitions, the wavelength of the transitions in superfluid helium under
  saturated vapour pressure and the pressure line shifts. Also mentioned is
  the change of the wavelength because
  of a pressure increase relative to the wavelength of the transitions at
  saturated vapour pressure.}
\end{table*}
The pressure dependence of the three emission lines is:
\begin{itemize}
\item $3s4s^3S_1 \rightarrow 3s3p^3P_0\,$:\\
$\lambda=(517.11\pm0.04)\,nm-(0.09\pm0.01)\,\frac{nm}{bar}\cdot p\, [bar]$
\item $3s4s^3S_1 \rightarrow 3s3p^3P_1\,$:\\
$\lambda=(517.51\pm0.06)\,nm-(0.06\pm0.01)\,\frac{nm}{bar}\cdot p\,
  [bar]$
\item $3s4s^3S_1 \rightarrow 3s3p^3P_2\,$:\\
$\lambda=(518.52\pm0.04)\,nm-(0.06\pm0.01)\,\frac{nm}{bar}\cdot p\, [bar]$
\end{itemize}
The deviation in wavelength between the theoretical and the
experimental curves is due to the precision of the pair potentials and is
rather small compared to other
calculations \cite{mbook}, e.g. for barium atoms in superfluid helium the
deviation is about $14\,$nm \cite{mdiss}.\\
The quality of the agreement of the calculated and measured pressure shifts
for all three lines can be tested with a
statistical hypothesis test, the students test. The deviation of the three 
 values is compatible with statistical fluctuations. 
Therefore a mean pressure line
shift of ($0.07 \pm 0.01\,$nm$/$bar) can be derived. This
very good consistency between the experimental and the theoretical values
allows the conclusion that the magnesium
atom seem to  maintain a bubble like structure under 
increased helium pressures. The pressure shift is monotonous.  
\section{Discussion}
\label{sec:dis}
As a consequence of the higher pressure the
bubble like defect shrinks, i.e. the equilibrium radius
decreases. The repulsive part of the pair potential energies 
due to Pauli forces rises in the upper P state already at larger radii than for the 
lower S state which implies a smaller wavelength for emitted radiation.\\  
Up to now only few pressure dependent measurements of electronic transitions
of foreign particles implanted into superfluid helium exist (see table
\ref{tab:over}). A quantitative comparison between the published line shifts
and the results presented in this paper is not possible for the line shift
themselves, because different types of transitions have been investigated.
Since the foreign atom-helium interactions potentials are not
comparable with each other, 
the different shifts for the various elements are not surprising.
Interesting is a comparison concerning the relative
pressure shift in  wavelength which is much larger for the electron
bubble than for any other structure. This reflects the fact that the electron
bubble is much more compressible than the other bubbles.
The similarity of the relative line shifts, i.e. the change of
wavelength with pressure relative to the transition wavelength at saturated
vapor pressure, for Mg, Rb, Ba, Tm and He$_2$ may be taken as indication that in
all these cases bubbles are formed with similar size and compressibility.
The within statistics linear behaviour of the pressure shifts suggests
smooth and continuous change in the size and structure of the defect caused by 
all these systems.
\begin{acknowledgement}
This work was supported in part by the Deutsche Forschungsgemeinschaft
(DFG). We would like to express our thanks to B. Tabbert and M. Foerste for
their input at an early stage and their constant interest and
suggestions. A. K. would like to acknowledge an Alexander von Humboldt
postdoctoral fellowship. 
\end{acknowledgement}

%
%
%

\begin{thebibliography}{}
\bibitem{henshaw}
D.G. Henshaw, A.D. Woods, Phys. Rev. \textbf{121}, 1961, 1266-1274 
\bibitem{toennies} J.P. Toennies and A.F. Vilesov, Ann. Rev. Phys. Chem. 
\textbf{49}, 1998, 1-40; and references therein
\bibitem{weis}S.I. Kanorsky and A. Weis, Adv. in Atomic, Molecular and Optical Physics
\textbf{38}, 1998, 87-120; and references therein
\bibitem{berndrev}
B. Tabbert, H. G\"unther, G. zu Putlitz, J.Low Temp. Phys. \textbf{5/6}, 1997, 653-707;
and references therein
\bibitem{atkins}
K. R. Atkins, Phys. Rev. \textbf{116}, 1959, 1339-1343
\bibitem{mgpap}
H. G\"unther, M. Foerste, C. H\"onninger, G. zu Putlitz, B. Tabbert, Z. Phys. B \textbf{98}, 1995, 395-398. 
\bibitem{bubb1}
A.P. Hickman, W. Steets, N.F. Lane, Phys. Rev. B \textbf{12}, 1975, 3705-3717
\bibitem{inapap}
I. Baumann, M. Foerste, K. Layer, G. zu Putlitz, B. Tabbert, Ch. Z\"uhlke,
J. Low. Temp. Phys. \textbf{110}, 1998, 213-218
\bibitem{bubb2}
A.P. Hickman, N.F. Lane, Phys. Rev. Lett. \textbf{26}, 1971, 1216-1219
\bibitem{ebubb} 
C.C. Grimes, G. Adams, Phys. Rev. B \textbf{41}, 1990, 6366-6371
\bibitem{Donn}
R.J. Donnelly, \textit{Experimental Superfluidity}, (W.I. Glaberson and P.E. 
Parks (eds.) University of Chicago Press, Chicago, 1967), 226-227
\bibitem{beau}
M. Beau, Dissertation, University Heidelberg (1990)
\bibitem{pair}
Q. Hui, Ph.D. Thesis, Saitama University (1997)
\bibitem{mbook}
M. Foerste, I. Baumann, U. Pritzsche, G. zu Putlitz, B. Tabbert, J.
 Wiebe, C. Z\"uhlke, Optical and mobility measurements of
  alkali earth atoms and ions in superfluid helium, \textit{Advances in solid
    state physics}, (B. Kramer, Friedr. Vieweg \& Sohn,
  Braunschweig/Wiesbaden, 1999) 355-367
\bibitem{mdiss}
M. Foerste, Dissertation, University Heidelberg (1997)
\bibitem{ref:grim92}
C.C. Grimes, G. Adams, Phys. Rev. B \textbf{45}, 1992, 2305-2310
\bibitem{ref:golov95}
A. Golov, Z. Phys. B \textbf{98}, 1995, 363-366
\bibitem{ref:hill}
J.C. Hill, O. Heybey, G.K. Walters, Phys. Rev. Lett. \textbf{26}, 1971, 1213-1216
\bibitem{ref:soley}
F.J. Soley, W.A. Fitzsimmons, Phys. Rev. Lett. \textbf{32}, 1974, 988-991
\bibitem{ref:taka93}
Y. Takahashi, K. Sano, T. Kinoshita, T. Yabuzaki,
Phys. Rev. Lett. \textbf{71}, 1993, 1035-1038
\bibitem{ref:kino951}
T. Kinoshita, K. Fukada, Y. Takahashi, T. Yabuzaki, Phys. Rev. A \textbf{52},
1995, 2707-2716
\bibitem{ref:bauer901}
H. Bauer, M. Beau, B. Friedl, C. Marchand, K. Miltner, H.J. Reyher, 
Phys. Lett A \textbf{146}, 1990, 134-140
\bibitem{ref:kan951}
S. Kanorsky, A. Weis, M. Arndt, R. Dziewior, T.W. H\"ansch, Z. Phys. B
\textbf{98}, 1995, 371-376
\bibitem{ref:kino952}
T. Kinoshita, K. Fukada, T. Yabuzaki, Z. Phys. B \textbf{98}, 1995, 387-390
\bibitem{ref:ishika}
K. Ishikawa, A. Hatakeyama, K. Gosyono-o, S. Wada, Y. Takahashi, T. Yabuzaki,
Phys. Rev. B \textbf{56}, 1997, 780-787
%
\end{thebibliography}
%

\end{document}